\documentclass[superscriptaddress,secnumarabic,amssymb,amsmath,nobibnotes,aps,prd,showkeys,showpacs,twocolumn,nofootinbib]{revtex4-1}%

\usepackage{hyperref}
\usepackage{amsmath}
\usepackage{amssymb}
\usepackage{graphicx}
\usepackage{color}
\usepackage{subfig}
\usepackage{bm}
\usepackage{gensymb}
\usepackage{bigints}
\usepackage[usenames,dvipsnames,svgnames,table]{xcolor}
\usepackage{pifont}
\usepackage{aas_macros}
\usepackage{orcidlink}
\usepackage{physics}
\graphicspath{ {./images/} }
\usepackage[normalem]{ulem}

\renewcommand{\d}{\mathrm{d}}
\newcommand{\e}{\operatorname{e}}

\usepackage[labelfont=bf]{caption}
\hypersetup{
    colorlinks=true,
    linkcolor=Blue,
    filecolor=Blue,
    urlcolor=MidnightBlue,
    citecolor=Blue,
   pdftitle={Cooling age}
}

\begin{document}


\title{Crystallized white dwarf stars in scalar-tensor gravity}


\author{Sof\'ia Vidal\orcidlink{0000-0003-0143-0427}}
\thanks{Corresponding author}
\email{sofia.vidal@ut.ee}
\affiliation{Laboratory of Theoretical Physics, Institute of Physics, University of Tartu,
W. Ostwaldi 1, 50411 Tartu, Estonia
}

\author{Aneta Wojnar\orcidlink{0000-0002-1545-1483}}
\email[E-mail: ]{awojnar@ucm.es}
\affiliation{Department of Theoretical Physics \& IPARCOS, Complutense University of Madrid, E-28040, 
Madrid, Spain 
}

\author{Laur J\"arv\orcidlink{0000-0001-8879-3890}}
\email[E-mail: ]{laur.jarv@ut.ee}
\affiliation{Laboratory of Theoretical Physics, Institute of Physics, University of Tartu,
W. Ostwaldi 1, 50411 Tartu, Estonia
}

\author{Daniela Doneva\orcidlink{0000-0001-6519-000X}}
\email[E-mail: ]{daniela.doneva@uni-tuebingen.de}
\affiliation{Theoretical Astrophysics, Eberhard Karls University of T\"ubingen, T\"ubingen 72076, Germany}
\affiliation{INRNE - Bulgarian Academy of Sciences, 1784 Sofia, Bulgaria}

\begin{abstract}
We study the effects of massive scalar-tensor theories on the internal properties, crystallization, and cooling process of white dwarf stars that might potentially solve observational tensions. We show that these modified gravity theories alter the inner structure of the star leading to sub-Chandrasekhar mass white dwarfs. This further results in a modification of Debye temperature, ion and electron specific heats. Finally, we find that the cooling process is significantly shortened in scalar-tensor theories leading to reduced cooling ages.
\end{abstract}

\maketitle

\section{Introduction}

White dwarfs (WDs) represent the final evolutionary stage of stars that originally had masses less than about 10 solar masses \cite{1986bhwd.book.....S,2018MNRAS.480.1547L}. These stars, having exhausted their nuclear fuel, leave behind dense cores primarily composed of elements such as carbon and oxygen. 
Although WDs with masses below approximately 0.4 solar masses tend to be primarily composed of helium, other white dwarfs generally contain heavier elements like carbon and oxygen. Additionally, in all these objects, the core is encased in a thin layer of helium, which is further enveloped by an even thinner layer of hydrogen.
Unlike main sequence stars, WDs do not generate energy through thermonuclear reactions; instead, their evolution is primarily governed by cooling processes. The energy reservoir of a WD is its degenerate core, which releases energy through the outer non-degenerate layers. Simple models often assume that WDs have temperatures well below the Fermi temperature, allowing for zero-temperature calculations to describe their structure. However, temperature is crucial for determining their ages. By analyzing cooling processes, we can estimate the lifetimes of these stars, providing valuable insights into the age of various galactic components, the stellar formation rate, and the history of the galaxy.

The structure of a WD is supported by electron degeneracy pressure, a quantum mechanical effect arising from the Pauli exclusion principle, which counterbalances the inward pull of gravity. This unique balance allows WDs to remain stable unless their mass approaches the Anderson--Stoner--Chandrasekhar limit \cite{1929ZPhy...56..851A,1930LEDPM...9..944S,1931ApJ....74...81C}, approximately 1.4 solar masses for non-rotating, non-magnetized white dwarfs. Exceeding this limit typically leads to a type Ia supernova, a critical event used as standard candles in astronomy due to their consistent peak luminosities \cite{choudhuri2010astrophysics,1987ApJ...323..140L,1997Sci...276.1378N}.

Recent observations, however, have revealed peculiar over-luminous and under-luminous type Ia supernovae, suggesting that the traditional Chandrasekhar limit might not be universally applicable. Over-luminous supernovae \cite{2006Natur.443..308H}, for example, may indicate the presence of super-Chandrasekhar white dwarfs \cite{2012ApJ...756..191K}, while under-luminous ones \cite{1992AJ....104.1543F} suggest the existence of sub-Chandrasekhar white dwarfs \cite{2006A&A...460..793S}. These anomalies challenge existing theories and hint at the need for modified gravity theories to explain the observed phenomena.

Surveys such as GAIA, SDSS, and Kepler have significantly advanced our understanding of WDs by providing detailed data on their surface temperatures and masses. These studies have shown that most WDs have masses below 1 solar mass, with no detections of super-Chandrasekhar white dwarfs, which could be explained by their potential high magnetization leading to reduced luminosity or instability due to extremely strong magnetic fields. Additionally, recent observations from the GAIA satellite have confirmed that the cores of cooling white dwarf stars undergo the process of crystallization \cite{2019NatAs...3..129V}.

Moreover, WDs in binary systems or with non-spherical shapes can emit gravitational waves (GWs). These waves are generated if the system has a non-zero quadrupole moment. Continuous gravitational waves ({emitted at certain frequency and amplitude}) can be {also generated} by a single spinning massive WD with axisymmetry breaking, such as through the misalignment of the magnetic field and the rotation axes or the presence of surface ``mountains'' \cite{1979PhRvD..20..351Z}.  
It has been suggested that future space-based gravitational wave detectors like LISA, DECIGO, or BBO could detect the gravitational radiation emitted by such an isolated magnetized white dwarf pulsar in the future \cite{2019MNRAS.490.2692K,2023mgm..conf.4461S,2017PhRvD..95h4029B}. Also, depending on the field geometry
and its strength, those detectors can detect this gravitational radiation for a long time, allowing to
estimate the size of the super-Chandrasekhar WDs,
where the electromagnetic surveys are not successful \cite{2020ApJ...896...69K}. This can provide additional insight into the characteristics and behaviors of white dwarfs, contributing further to our understanding of their properties and the fundamental physics governing their existence.

In recent years, scalar-tensor theories (STT) have garnered significant attention in the literature as a prototypical extension of general relativity \cite{Brans:1961sx,Damour:1992we,CANTATA:2021asi}. They arise naturally from compact higher dimensions, by taking into account quantum corrections to a scalar field minimally coupled
to gravity, or in endeavors to build a theory that is fundamentally scale-free. In cosmology many of the phenomenologically successful models of early universe inflation or late universe dark energy belong to the STT family \cite{CANTATA:2021asi}. While in astrophysics much of the existing research has focused on neutron stars within various STT models (for review, see \cite{Olmo:2019flu,2024RvMP...96a5004D}), white dwarfs (WDs) have also been studied within this framework \cite{2018JCAP...05..028S,2020PhRvD.101b3017P, 2012MNRAS.423.3328F,2016PhRvL.116o1103J,2022PhRvD.106l4010A,2019PhRvD.100b4025C,2019CQGra..36i5017L,1999MNRAS.305..905B}.

In the following sections, we will examine the cooling process of WDs within the framework of STT using the full relativistic theory. Although white dwarfs are typically large and often studied within the non-relativistic regime of general relativity (GR) or modified gravity (MG), it has been shown that relativistic effects can significantly impact certain phenomena \cite{Olmo:2019flu,2021JCAP...02..026M,2021PhRvD.104l3005H,2023PhRvD.107d4025W}. We start by introducing the hydrostatic equilibrium equations in STT and GR in Sec.\ \ref{sec:hydrostaticequil} as well as the Chandrasekhar equation of state (EoS). We outline the surface structure and processes relevant to the cooling of a WD in Sec.\ \ref{sec:WDinSTT}, followed by the corresponding equations for GR in Sec.\ \ref{sec:WDinGR}. In Sec.\ \ref{sec:numerics}, we provide a discussion on the numerics as well as our numerical results. We demonstrate that internal properties of WDs are modified by the considered gravity theories and thus, lead to an altered cooling age. Finally, we conclude in Sec.\ \ref{sec:conclusions}.

\section{Hydrostatic Equilibrium Equations} \label{sec:hydrostaticequil}

{Among the several possible reparametrizations of STT, it is natural to assume that the physical units are defined in the so-called Jordan frame, where the free particles follow geodesics and the matter fields are covariantly conserved \cite{Damour:1992we,Jarv:2014hma}. However, for the sake of} 
mathematical convenience, {one often invokes the Einstein frame, where the scalar field is clearly disentangled from the metric tensor, and the gravitational constant is truly constant. In the current paper we will also employ the Einstein frame} in most of the intermediate calculations. It is related to the physical Jordan frame through a conformal transformation of the metric and scalar field redefinition, {whereas} for a detailed discussion of the two frames we refer the reader to Refs.\ \cite{Damour:1992we,Doneva:2013qva,2024RvMP...96a5004D,Jarv:2014hma}. To avoid confusion, the quantities in the Einstein frame will be denoted with a tilde, e.g.\ $\Tilde{r}$, while the ones without, e.g.\ $r$, are in the Jordan frame. 

\par In the following subsections, we will discuss the equations describing stellar equilibrium in GR and scalar-tensor gravity.

\subsection{Scalar-Tensor Theory}

\par The action of scalar-tensor theories in the Einstein frame reads \cite{Damour:1992we}
\begin{align}\label{eq:action}
    S =& \frac{{c^4}}{16\pi G}\int \frac{\d^4x}{{c}} \sqrt{-g} \left[\Tilde{R} - 2\Tilde{g}^{\mu\nu}\tilde\partial_{\mu}\Tilde{\varphi}\tilde\partial_{\nu}\Tilde{\varphi}  - V(\Tilde{\varphi})\right] \nonumber \\
    &+ S_\mathrm{m}\left(A^2(\Tilde{\varphi})\Tilde{g}_{\mu\nu}, \chi\right)\;,
\end{align}
where $G$ denotes the bare gravitational constant and $V(\Tilde{\varphi})$ is the potential of the scalar field  $\Tilde{\varphi}$. The conformal factor $A(\Tilde{\varphi}) >0$ connects the Einstein frame and the physical Jordan frame metric through the {rescaling} transformation $g_{\mu\nu} = A^2(\Tilde{\varphi})\Tilde{g}_{\mu\nu}$.
Varying the action \eqref{eq:action} with respect to the metric and scalar field, one can obtain the following set of equations
\begin{align}
    \Tilde{R}_{\mu\nu} - \frac{1}{2}\Tilde{g}_{\mu\nu}\Tilde{R} =&\; \frac{8\pi G}{c^4} \mathcal{\Tilde{T}}_{\mu\nu} + 2\tilde\nabla_{\mu}\Tilde{\varphi}\tilde\nabla_{\nu}\Tilde{\varphi} \nonumber \\
    &-g_{\mu\nu}g^{\alpha\beta}\tilde\nabla_{\alpha}\Tilde{\varphi}\tilde\nabla_{\beta}\Tilde{\varphi}-\frac{1}{2}V(\Tilde{\varphi})\Tilde{g}_{\mu\nu}\;, \\
    \tilde\nabla_{\mu}\tilde\nabla^{\mu}\Tilde{\varphi} =& -\frac{4\pi G}{{c^4}}\alpha(\Tilde{\varphi})\mathcal{\Tilde{T}} + \frac{1}{4}\frac{\d V(\Tilde{\varphi})}{\d\Tilde{\varphi}} \,,
\end{align}
where $\alpha(\Tilde{\varphi}) = \frac{\d\ln{A(\Tilde{\varphi})}}{\d\Tilde{\varphi}}$.
The energy-momentum tensor $\mathcal{\tilde T_{\mu\nu}}$ is the one of the perfect fluid, with the properly transformed mass density and pressure, that is, $ \Tilde{\rho} = A^4(\Tilde{\varphi})\rho$ and $\Tilde{p} = A^4(\Tilde{\varphi})p$, where $p$ and $\rho$ are the (physical) thermodynamic variables related by an equation of state.

{In the context of STT reparametrizations, the quantity $\alpha(\Tilde{\varphi})$ has an invariant meaning as a measure of the coupling between the scalar field and the metric \cite{Damour:1992we,Jarv:2014hma}.
Vanishing $\alpha(\Tilde{\varphi})$ corresponds to a minimal coupling, i.e.\  general relativity with a scalar field, while nonvanishing $\alpha(\Tilde{\varphi})$ indicates nonminimal coupling whereby the scalar field can be understood to act as an extra mediator of gravity, sourced by the trace of the matter energy-momentum tensor.}
In what follows, we will focus on a massive Brans-Dicke theory {\cite{Brans:1961sx,Damour:1992we}} defined by a constant $\alpha_0$ and scalar field mass $m_{\Tilde{\varphi}}$ through
\begin{equation}
    \alpha(\Tilde{\varphi}) = \alpha_0 \quad\mathrm{and}\quad V(\Tilde{\varphi}) = \frac{2m^2_{\Tilde{\varphi}}c^2}{\hbar^2}\Tilde{\varphi}^2\;.
    \label{eq:paramspotentialSTT}
\end{equation}
{The mass $m_{\tilde{\varphi}}$ characterizes the effective range of the scalar force, the more massive the sooner the extra scalar field induced Yukawa correction to the Newtonian gravitational potential will die off in distance \cite{Jarv:2014hma}.}

We assume a spherically symmetric, static metric
\begin{equation}\label{eq:metricSTT}
    \d \Tilde{s}^2 = -\e^{2\Tilde{\phi}(r)}c^2\d t^2 + \e^{2\Tilde{\Lambda}(r)}\d \Tilde{r}^2 + \Tilde{r}^2\d\Omega^2\;,
\end{equation}
with the angular element ${\Omega^2 = \d\theta^2 + \sin^2{\theta}\d\vartheta^2}$.
Thus, the Einstein frame field equations describing the interior structure of a WD in a static and spherically symmetric configuration reduce to
{\small{\begin{subequations}
\label{eq:scalartensorFE}
    \begin{align}
        \frac{2\Tilde{\Lambda}'}{\Tilde{r}} &= \frac{8\pi G}{c^2} A^4(\Tilde{\varphi})\rho\e^{2\Tilde{\Lambda}} + \frac{1-\e^{2\Tilde{\Lambda}}}{\Tilde{r}^2} +\Tilde{\varphi}'^2 +\frac{V(\Tilde{\varphi})}{2}\e^{2\Tilde{\Lambda}}\;, \label{eq:lambdaSTTFE} \\
        \frac{2\Tilde{\phi}'}{\Tilde{r}} &= \frac{8\pi G}{c^4} A^4(\Tilde{\varphi})p\e^{2\Tilde{\Lambda}} - \frac{1-\e^{2\Tilde{\Lambda}}}{\Tilde{r}^2} +\Tilde{\varphi}'^2-\frac{V(\Tilde{\varphi})}{2}\e^{2\Tilde{\Lambda}}\;, \label{eq:phiSTTFE} \\
        \Tilde{\varphi}'' &+ \left(\Tilde{\phi}'-\Tilde{\Lambda}'+\frac{2}{\Tilde{r}}\right)\Tilde{\varphi}' \nonumber\\
        &\quad\quad\quad= \left[\frac{4\pi G}{{c^4}}\alpha A^4(\Tilde{\varphi})(\rho c^2-3p) + \frac{1}{4}\frac{\d V}{\d\Tilde{\varphi}}\right]\e^{2\Tilde{\Lambda}}\;, 
    \end{align}
\end{subequations}}}
where primes denote derivatives with respect to the radius $\Tilde{r}$. 
These equations have to be supplemented by the equation for hydrostatic equilibrium
\begin{equation}
    p' = -(\rho c^2 + p)\left(\Tilde{\phi}'+\alpha\Tilde{\varphi}' \right)\;.\label{eq:hydrostaticequilibriumSTT}
\end{equation}
\par Note that in the equations above the hydrodynamical quantities, such as the density and the pressure, are written without a tilde. Thus, they are in the physical Jordan frame. This is done for convenience because the equation of state, discussed below, is also naturally provided in the physical frame.
\par The boundary conditions are given at the center of the star by
\begin{equation*}
    p(\Tilde{r}=0) = p_c \quad\mathrm{and}\quad \Tilde{\Lambda}(\Tilde{r}=0) = 0\;,
\end{equation*}
where the central pressure $p_c$ is a free input parameter. The spatial asymptotic of the metric functions and the scalar field at infinity is given by
\begin{equation*}
    \lim_{\Tilde{r}\to\infty} \Tilde{\Lambda} (\Tilde{r}) = 0 , \;\;\lim_{\Tilde{r}\to\infty} \Tilde{\phi}(\Tilde{r}) = 0 \quad\mathrm{and}\quad \lim_{\Tilde{r}\to\infty} \Tilde{\varphi}(\Tilde{r}) = 0\;.
\end{equation*}

\par We define the stellar radius $\Tilde{r}_s$ in Einstein frame as the radius for which $p(\Tilde{r}_s) = 0$. The circumferential radius of the star, in the physical Jordan frame, is given by $R_s = A[\Tilde{\varphi}(\Tilde{r}_s)]\Tilde{r}_s$.

\subsection{General Relativity}

We can obtain the {plain} general relativistic case by setting {$\alpha(\Tilde{\varphi})=0$, and letting the scalar field vanish identically,} $\Tilde{\varphi}(r) = 0$. 
{It is also convenient to assume $A(\Tilde{\varphi})=1$ to exactly match the units, e.g.\ $\tilde{r}=r$.}
The hydrostatic equilibrium equations {\eqref{eq:scalartensorFE}} reduce then to
\begin{subequations}
    \begin{align}
        \Lambda' &= \frac{4\pi r G}{c^2} \rho\e^{2\Lambda} + \frac{1-\e^{2\Lambda}}{2r}\;, \\
        \phi' &= \frac{4\pi r G}{c^4} p\e^{2\Lambda} - \frac{1-\e^{2\Lambda}}{2r}\;, \\
        p' &= -(\rho c^2 + p)\;\phi'\;. \label{eq:hydrostaticequilibriumGR}
    \end{align}
    \label{eq:generalrelativityFE}
\end{subequations}
\par Note that now we are dealing with the spherical-symmetric metric of the form
\begin{equation}\label{eq:metricGR}
    \d s^2 = -\e^{2\phi(r)}c^2\d t^2 + \e^{2\Lambda(r)}\d r^2 + r^2\d\Omega^2\;,
\end{equation}
which in the case of the scalar-tensor model would be also the one of the Jordan frame.

\subsection{Equation of state}
The systems of differential equations, \eqref{eq:scalartensorFE} and \eqref{eq:generalrelativityFE}, {can be} closed with an appropriate equation of state (EoS). We consider the Chandrasekhar EoS describing a fully degenerate relativistic gas. In this case, defining the constants
\begin{equation*}
    B_1 := \frac{\pi m_e^4 c^5}{3h^3}\;, \quad B_2:= \frac{8\pi \mu_e m_p (m_e c)^3}{3h^3}\;,
\end{equation*}
the pressure and mass density are given respectively by \cite{Glendenning:1997wn}
\begin{subequations}
 \label{eq:ChandrasekharEOS}
    \begin{align}
        p(x) &= B_1 \left[(2x^3-3x)\sqrt{1 + x^2} + 3\sinh^{-1}{x} \right]\;, \label{eq:ChandrasekharPressure} \\
        \rho (x) &= \rho_0 (x) + \rho_e (x) \label{eq:ChandrasekharDensity} \\
        &= B_2x^3 + \frac{9B_1}{c^2} \left[(2x^3+x)\sqrt{1 + x^2} - \sinh^{-1}{x} \right]\;, \nonumber
    \end{align}
\end{subequations}
where $x=p_F/m_ec$ represents the dimensionless Fermi momentum, $p_F$ the dimensionful Fermi momentum, $m_e$ the electron mass, $h$ Planck's constant, $\mu_e$ the mean molecular weight per electron, and $m_p$ the proton mass. The first term in \eqref{eq:ChandrasekharDensity} defines the rest mass density, while the latter corresponds to the kinetic contribution.

\section{Crystallization of white dwarfs in scalar-tensor gravity} \label{sec:WDinSTT}

Following the analysis in \cite{2023PhRvD.107d4072K} for Palatini $f(R)$ gravity, we now present a simplified cooling model for WDs in scalar-tensor theories. This formalism takes not only the residual thermal energy of the star into account, but also the latent heat that is released during the crystallization process of the core. This energy source contributes significantly as it will become evident in the resulting cooling ages presented below. Analogous considerations in the framework of non-relativistic physics can be found in \cite{1986bhwd.book.....S}.
\par We will first derive the luminosity that arises from the structure of the surface layers. This quantity is needed to get the cooling age of the WD. Then, we continue to describe the energy loss through the decreasing thermal energy and the effects of crystallization during this process. Lastly, we obtain the cooling age of a WD in STT.

\subsection{Surface Layer Structure} \label{sec:surfaceSTT}
The interior of a WD consists of completely degenerate matter, but is surrounded by nondegenerate surface layers. Close to the surface, the diffusion of photons leads to an outward energy flux \cite{1995PhT....48i..94H} 
\begin{equation}
    L = -4\pi r^2 \frac{c}{3\kappa\rho} \frac{\d}{\d r} \left(aT^4\right)\;,
    \label{eq:photondiffusion}
\end{equation}
where $L$ is the luminosity, $\kappa$ the opacity, and $aT^4$ the black body radiation with $a = 8\pi^5k_B^4 / 15c^3h^3 \approx 7.6\times 10^{-15}\; \mathrm{erg} \, \mathrm{cm}^{-3} \mathrm{K}^{-4}$ the radiation constant.
\par On the other side, writing $\e^{2\Tilde{\Lambda}}=\left(1-\frac{2Gm}{\Tilde{r}c^2}\right)^{-1}$ as well as using \eqref{eq:phiSTTFE} and \eqref{eq:hydrostaticequilibriumSTT}, we can express the hydrostatic equilibrium equation as
{\small{\begin{align}
    p' =& -\frac{Gm}{\Tilde{r}^2c^2}\left(\rho c^2+p\right)\left[1 + \frac{4\pi \Tilde{r}^3}{m}\left(\frac{A^4(\Tilde{\varphi})p}{c^2} + \frac{V(\Tilde{\varphi})c^2}{16\pi G}\right)\right] \nonumber \\
    &\quad\quad\times\left(1-\frac{2Gm}{\Tilde{r}c^2}\right)^{-1} -(\rho c^2 + p)\left(\alpha\Tilde{\varphi}' + \frac{\Tilde{r}\Tilde{\varphi}'^2}{2}\right) \nonumber \\
    \approx&-\frac{Gm\rho}{\Tilde{r}^2}\left(1+\frac{p}{\rho c^2}\right)\left(1 + \frac{\Tilde{r}^3c^2}{4Gm}V(\Tilde{\varphi})\right)\left(1-\frac{2Gm}{\Tilde{r}c^2}\right)^{-1} \nonumber \\
    & -(\rho c^2 + p)\left(\alpha\Tilde{\varphi}' + \frac{\Tilde{r}\Tilde{\varphi}'^2}{2}\right)\;. \label{eq:hydrostaticequilibriumSTTextended}
\end{align}}}
An order of magnitude estimate for WD reveals that ${4\pi\Tilde{r}^3p/c^2 \ll m}$ close to the surface and therefore the corresponding term can be dropped in the last step. 
\par The temperature gradient is given by \cite{1995PhT....48i..94H}
\begin{equation}
    T' = -\frac{3L\kappa\Tilde{\rho}}{4\pi \Tilde{r}^2 4 acT^3} = -\frac{L\kappa A^4(\Tilde{\varphi})\rho}{4\pi \Tilde{r}^2 4 acT^3}\;,
    \label{eq:TempGradientSTT}
\end{equation}
Dividing the hydrostatic equilibrium equation \eqref{eq:hydrostaticequilibriumSTTextended} by \eqref{eq:photondiffusion} and using Kramer's opacity ${\kappa = \kappa_0 \rho T^{-3.5}}$ results in
{\small{\begin{align}
    \frac{\d p}{\d T} =& \frac{4\pi Gm\; 4ac}{3L A^4(\Tilde{\varphi})\rho\kappa_0}\Biggl\{\Tilde{r}^2c^2\left(1+\frac{p}{\rho c^2}\right)\left(\alpha\Tilde{\varphi}' + \frac{\Tilde{r}\Tilde{\varphi}'^2}{2}\right) \Biggr.\nonumber \\
    +& \Biggl.\left(1+\frac{p}{\rho c^2}\right)\left(1 + \frac{\Tilde{r}^3c^2}{4Gm}V(\Tilde{\varphi})\right)\left(1-\frac{2Gm}{\Tilde{r}c^2}\right)^{-1}\Biggr\} T^{6.5}\;.
\end{align}}}
\par Close to the surface, we can use the equation of state for nondegenerate matter
\begin{equation}
    \rho = \frac{\mu m_u}{k_BT}p \quad\rightarrow\quad \frac{p}{\rho c^2} = \frac{k_BT}{\mu m_uc^2}\;,
    \label{eq:nondegEOS}
\end{equation}
with the atomic mass $m_u$ and mean molecular weight $\mu$. An order of magnitude estimate further shows that
\begin{equation*}
    \frac{k_BT}{\mu m_u c^2} \approx 6.7 \times 10^{-14} K^{-1} T \ll 1\;,
\end{equation*}
since $T[K] \in [10^6, 10^8]$. Thus, the term $p/\rho c^2$ can be safely neglected in further considerations. Using \eqref{eq:nondegEOS} to replace the remaining energy density with pressure, we arrive at
\begin{align}
    \frac{\d p}{\d T} =& \frac{4\pi Gm\; 4ac}{3L A^4(\Tilde{\varphi})\kappa_0}\frac{k_\mathrm{B}}{\mu m_u}\Biggl\{\Tilde{r}^2c^2\left(\alpha\Tilde{\varphi}' + \frac{\Tilde{r}\Tilde{\varphi}'^2}{2}\right) \Biggr.\nonumber \\
    &+ \Biggl.\left(1 + \frac{\Tilde{r}^3c^2}{4Gm}V(\Tilde{\varphi})\right)\left(1-\frac{2Gm}{\Tilde{r}c^2}\right)^{-1}\Biggr\} \frac{T^{6.5}}{p}\;.
    \label{eq:dpdTSTT}
\end{align}
Since the surface layer is thin compared to the stellar radius, we can assume $\Tilde{r}\approx\Tilde{r}_s$. Accordingly, we approximate all quantities but the temperature on the RHS of \eqref{eq:dpdTSTT} to their value at $\Tilde{r}_s$, neglecting small variations. Then, integrating the resulting equation with respect to this temperature with the boundary condition $p(T=0)=0$ yields
\footnote{Note that the core temperature is not zero, even for degenerate matter as considered here; however, in comparison to the surface temperature, the core temperature is much lower and as a result, in our toy-model considerations, we can take $T=0$.}
\begin{align}
    p^2 =& \frac{2}{7.5}\frac{4\pi GM\; 4ac}{3L A^4(\Tilde{\varphi}_s)\kappa_0}\frac{k_\mathrm{B}}{\mu m_u}\Biggl\{\Tilde{r}_s^2c^2\left(\alpha_s\Tilde{\varphi}'_s + \frac{\Tilde{r}_s\Tilde{\varphi}'^2_s}{2}\right) \Biggr.\nonumber \\
    &+ \Biggl.\left(1 + \frac{\Tilde{r}_s^3c^2}{4GM}V(\Tilde{\varphi}_s)\right)\left(1-\frac{2GM}{\Tilde{r}_sc^2}\right)^{-1}\Biggr\} T^{7.5}\;,
    \label{eq:surfacePressureSTT}
\end{align}
or respectively
\begin{align}
    \rho^2 =& \frac{2}{8.5}\frac{4\pi GM\; 4ac}{3L A^4(\Tilde{\varphi}_s)\kappa_0}\frac{k_\mathrm{B}}{\mu m_u}\Biggl\{\Tilde{r}_s^2c^2\left(\alpha_s\Tilde{\varphi}'_s + \frac{\Tilde{r}_s\Tilde{\varphi}'^2_s}{2}\right) \Biggr.\nonumber \\
    &+ \Biggl.\left(1 + \frac{\Tilde{r}_s^3c^2}{4GM}V(\Tilde{\varphi}_s)\right)\left(1-\frac{2GM}{\Tilde{r}_sc^2}\right)^{-1}\Biggr\}\; T^{8.5}\;,
    \label{eq:surfaceDensitySTT}
\end{align}
where the subscript $s$ denotes quantities evaluated at the stellar radius, e.g. $\Tilde{\varphi}_s = \Tilde{\varphi}(\Tilde{r}_s)$. So that we are left with a relation between pressure, or respectively density, and temperature while all other quantities are fixed.

The opacity is $\kappa_0 = 4.34 \times 10^{24}\; Z\left(1+X\right) \mathrm{cm}^5 \mathrm{K}^{3.5} \mathrm{g}^{-2}$, where $X$ denotes the hydrogen mass fraction and $Z$ the mass fraction of heavier elements \cite{1986bhwd.book.....S}.

This description holds until reaching the point below the surface where electrons become degenerate. By equating the nondegenerate and degenerate, non-relativistic electron pressure \cite{1986bhwd.book.....S}, we obtain an estimate of the density $\rho_*$ and temperature $T_*$ at this transition layer between degenerate and nondegenerate matter

\begin{equation}
    \frac{\rho_* k_B T_*}{\mu_e m_u} = (10^{13} \mathrm{cm}^4 \mathrm{s}^{-2} \mathrm{g}^{-2/3})\; \left(\frac{\rho_*}{\mu_e} \right)^{5/3}\;.
\end{equation}
Solving this equation for the transition layer density $\rho_*$, we can write ${\rho_*=\rho_0\mu_e T^{3/2}}$ where we have defined ${\rho_0:= 2.4\times10^{-8} \mathrm{g}\, \mathrm{cm}^{-3}\mathrm{K}^{-3/2}}$. Going back to \eqref{eq:surfaceDensitySTT}, we thus obtain a surface luminosity of
\begin{align}
    L_*^\mathrm{STT} =& \frac{2}{8.5}\frac{4\pi GM_\odot\; 4ac}{3 A^4(\Tilde{\varphi}_s)\rho_0^2\kappa_0}\frac{m_u}{k_\mathrm{B}}\frac{\mu}{\mu_e}\Biggl\{\frac{\Tilde{r}_s^2c^2}{GM}\left(\alpha\Tilde{\varphi}'_s + \frac{\Tilde{r}_s\Tilde{\varphi}'^2_s}{2}\right) \Biggr.\nonumber \\
    &+ \Biggl.\left(1 + \frac{\Tilde{r}_s^3c^2}{4GM}V(\Tilde{\varphi}_s)\right)\left(1-\frac{2GM}{\Tilde{r}_sc^2}\right)^{-1}\Biggr\} \frac{M}{M_\odot} T_*^{3.5} \nonumber \\
    :=&\; \frac{C}{A^4(\Tilde{\varphi}_s)}\; \Biggl\{\frac{\Tilde{r}_s^2c^2}{GM}\left(\alpha\Tilde{\varphi}'_s + \frac{\Tilde{r}_s\Tilde{\varphi}'^2_s}{2}\right) \Biggr.\nonumber \\
    &+ \Biggl.\left(1 + \frac{\Tilde{r}_s^3c^2}{4GM}V(\Tilde{\varphi}_s)\right)\left(1-\frac{2GM}{\Tilde{r}_sc^2}\right)^{-1}\Biggr\} \frac{M}{M_\odot} T_*^{3.5}\;,
    \label{eq:surfacelayerlumSTT}
\end{align}
where $C \approx 2\times 10^6\; \mathrm{erg} \mathrm{s}^{-1} \mathrm{K}^{-3.5}$ for a carbon-oxygen WD with $\mu_e = 2$, $\mu = 1.4$, and a surface composed of $90\%$ helium, $10\%$ heavier elements and no hydrogen. Recall that all quantities with subscript $s$ in \eqref{eq:surfacelayerlumSTT} are evaluated at the surface of the star.

\subsection{Thermal Heat} \label{subsec:thermalSTT}

One of the most significant sources of energy of a WD during its cooling process is the residual thermal energy from the progenitor star
\begin{equation*}
    U = \Bar{c}_v\frac{M}{A_a m_p}T\;,
\end{equation*}
where $A_a$ is the mean atomic weight, $m_p$ is the proton mass and the mean specific heat $\Bar{c}_v$ is
\begin{equation}
    \Bar{c}_v = \frac{1}{M}\int_0^M \d m \left(c_v^{\mathrm{el}} + c_v^{\mathrm{ion}}\right)\;.
\end{equation}
The luminosity given by the decrease of thermal energy is then
\begin{equation}
    L = -\frac{\d U}{\d t} = -\Bar{c}_v\frac{M}{A_a m_p}\frac{\d T}{\d t}\;.
    \label{eq:thermenergylum}
\end{equation}

The specific heat of the electrons per ion is
\begin{equation}
    c_v^{\mathrm{el}} = \frac{3}{2}\frac{k_B\pi^2}{3}Z\frac{k_BT}{\epsilon_F}\;,
    \label{eq:electronspecificheat}
\end{equation}
where $\epsilon_F$ is the Fermi momentum and $Z$ the charge. In comparison, the specific heat of the ions depends on the onset of crystallization. The ratio of Coulomb to thermal energy is
\begin{equation}
    \Gamma = 2.28\times 10^5 \,\frac{Z^2}{T}\left(\frac{\rho}{A_a}\right)^{1/3}\;,
    \label{eq:ratioGamma}
\end{equation}
so that crystallization starts at a critical point set by a specific $\Gamma_m$. Different computations adopt varying values of ${\Gamma_m\in\{60, 125\}}$ \cite{1966JChPh..45.2102B}. Throughout this work, we assume $\Gamma_m=60$ for illustrative purposes. Thus, denoting the Debye temperature by

\begin{equation}
    \Theta_D = 0.174\times10^4 \frac{2Z}{A_a}\sqrt{\rho}\;,
    \label{eq:DebyeTemp}
\end{equation}
and defining $y := \Theta_D/T$, $c_v^{\mathrm{ion}}$ is given by
\begin{subequations}
    \begin{align}
        \Gamma &< \Gamma_m: \quad c_v^{\mathrm{ion}} = \frac{3}{2}k_B\;, \\
        \Gamma &\geq \Gamma_m: \quad c_v^{\mathrm{ion}} = \frac{9k_B}{y^3}\int_0^{y}\d x \frac{x^4e^x}{\left(e^x-1\right)^2}\;.
    \end{align}
    \label{eq:ionspecificheat}
\end{subequations}

\subsection{Crystallization} \label{subsec:crystallSTT}

Crystallization is the liquid to solid phase transition of the matter at the center. It adds considerably to the energy loss of the star. Assuming the latent heat to be $qk_BT$, where $q$ is a numerical factor set to unity \cite{1968ApJ...151..227V}, the additional luminosity is
\begin{equation}
    L_q = qk_BT\frac{\d}{\d t}\left(\frac{m_s}{A_am_p}\right)\;,
    \label{eq:latentheat}
\end{equation}
where $m_s$ is the crystallized mass. Eq. \eqref{eq:latentheat} can be recast to
\begin{equation}
    L_q = 3\rho_sqk_B\frac{M}{A_am_p}\frac{1}{M}\left[\frac{\d m}{\d r}\frac{\d r}{\d\rho}\right]_{r=r_*}\frac{\d T}{\d t}\;,
    \label{eq:latentheatlum}
\end{equation}
where $\rho_s(T)$ is the crystallized mass density at temperature $T$ and $r_*$ is the radius for which $\rho(r_*) = \rho_s(T)$ holds.

\subsection{Cooling Age}

During the crystallization process, the total luminosity is obtained from Eqs.\ \eqref{eq:thermenergylum} and \eqref{eq:latentheatlum} as
\begin{equation}
    L_\mathrm{tot} = \frac{3k_BM}{A_am_p}\left(-\frac{\Bar{c}_v}{3k_B} + \frac{\rho_sq}{M}\left[\frac{\d m}{\d r}\frac{\d r}{\d\rho}\right]_{r=r_*}\right) \frac{\d T}{\d t}\;.
    \label{eq:totallum}
\end{equation}
Finally, combining Eqs.\ \eqref{eq:surfacelayerlumSTT} and \eqref{eq:totallum} we obtain a relation between temperature and time. We assume the cooling age of a WD to be the time it takes to cool down from an initial temperature of $T(t=0)=10^8$K to a present value of $10^6$K.

\section{Crystallization of white dwarfs in GR} \label{sec:WDinGR}

\subsection{Surface Layer Structure}

As mentioned above, we can obtain GR from 
scalar-tensor theory by setting $\alpha(\Tilde{\varphi})=0$, $\Tilde{\varphi}(r) = 0$. Hence, we can easily derive the surface layer equations in GR from our considerations in Sec.\ \ref{sec:surfaceSTT}. From \eqref{eq:dpdTSTT}, the derivative of the pressure with respect to the temperature becomes
\begin{equation}
    \frac{\d p}{\d T} = \frac{4\pi Gm\; 4ac}{3L\kappa_0}\frac{k_\mathrm{B}}{\mu m_u}\left(1-\frac{2Gm}{rc^2}\right)^{-1}\; \frac{T^{6.5}}{p}\;.
    \label{eq:dpdTGR}
\end{equation}

Consequently, the surface pressure depends on the temperature through
\begin{align}
    p^2 =& \frac{2}{7.5}\frac{4\pi GM\; 4ac}{3L\kappa_0}\frac{k_\mathrm{B}}{\mu m_u}\left(1-\frac{2GM}{R_sc^2}\right)^{-1} T^{7.5}\;,
    \label{eq:surfacePressureGR}
\end{align}
or, using the nondegenerate EOS \eqref{eq:nondegEOS}, respectively
\begin{align}
    \rho^2 =& \frac{2}{8.5}\frac{4\pi GM\; 4ac}{3L\kappa_0}\frac{k_\mathrm{B}}{\mu m_u}\left(1-\frac{2GM}{R_sc^2}\right)^{-1} T^{8.5}\;.
    \label{eq:surfaceDensityGR}
\end{align}
\par These equations lead to a surface luminosity of
\begin{align}
    L_*^\mathrm{GR} =& \frac{2}{8.5}\frac{4\pi GM_\odot\; 4ac}{3 \rho_0^2\kappa_0}\frac{m_u}{k_\mathrm{B}}\frac{\mu}{\mu_e}\left(1-\frac{2GM}{R_sc^2}\right)^{-1} \frac{M}{M_\odot} T_*^{3.5} \nonumber \\
    =&\; C\; \left(1-\frac{2GM}{R_sc^2}\right)^{-1} \frac{M}{M_\odot} T_*^{3.5}\;.
    \label{eq:surfacelayerlumGR}
\end{align}
with the constant $C$ and the chemical composition of the surface as defined above. To the authors' knowledge there are no studies on relativistic effects in stellar atmospheres.

Note that when considering Newtonian gravity, we can further assume $2GM/R_sc^2\ll 1$ and so the above equations reduce to
\begin{equation}
    \frac{\d p}{\d T} = \frac{4ac}{3} \frac{4\pi GM}{\kappa_0 L} \frac{T^{6.5}}{\rho}\;,
    \label{eq:dpdTNewt}
\end{equation}
and
\begin{equation}
    p = \left(\frac{2}{8.5}\frac{4ac}{3}\frac{4\pi GM}{\kappa_0 L} \frac{k_B}{\mu m_u} \right)^{1/2} T^{4.25}\;.
    \label{eq:surfacePressureNewt}
\end{equation}

In the end, the surface luminosity is simply given by
\begin{equation}
    L_*^\mathrm{Newt} = C \frac{M}{M_\odot} T_*^{3.5}\;.
    \label{eq:surfacelayerlumNewt}
\end{equation}

\subsection{Cooling Processes and Age}

The above considerations in Secs.\ \ref{subsec:thermalSTT} and \ref{subsec:crystallSTT} for the thermal heat and crystallization process are general and thus, also hold for GR. Modifications between the different theories enter the equations through the radial density, mass and Fermi momentum profiles that depend on the considered gravity theory. Here, we want to stress the following examples:
\begin{enumerate}
    \item The specific heat of the electrons per ion \eqref{eq:electronspecificheat} depends on the Fermi energy $\epsilon_F$
    \begin{equation*}
        \epsilon_F^2 = p_F^2c^2 + m_e^2c^4\;,
    \end{equation*}
    and the Fermi momentum $p_F$ is related to the density by \eqref{eq:ChandrasekharDensity}.
    \item The ratio of Coulomb to thermal energy \eqref{eq:ratioGamma} is subject to the density $\rho$. Thus, the critical value at which crystallization sets in will be reached at a different radius depending on the gravity theory.
    \item The density $\rho$ also enters the equation specifying the Debye temperature \eqref{eq:DebyeTemp}. Hence, the specific heat of ions, after the onset of crystallization, is altered.
    \item The additional luminosity \eqref{eq:latentheatlum} during crystallization varies according to the radius $r_*$, crystallized mass $\rho_s$ as well as radial mass and density profile. 
\end{enumerate}

All these influence the resulting cooling age for a WD obtained by now combining the equations \eqref{eq:surfacelayerlumGR}, or \eqref{eq:surfacelayerlumNewt} respectively, and \eqref{eq:totallum}.

\section{Numerical Results} \label{sec:numerics}

\begin{figure}[ht]
    \centering
    \includegraphics[width=\linewidth]{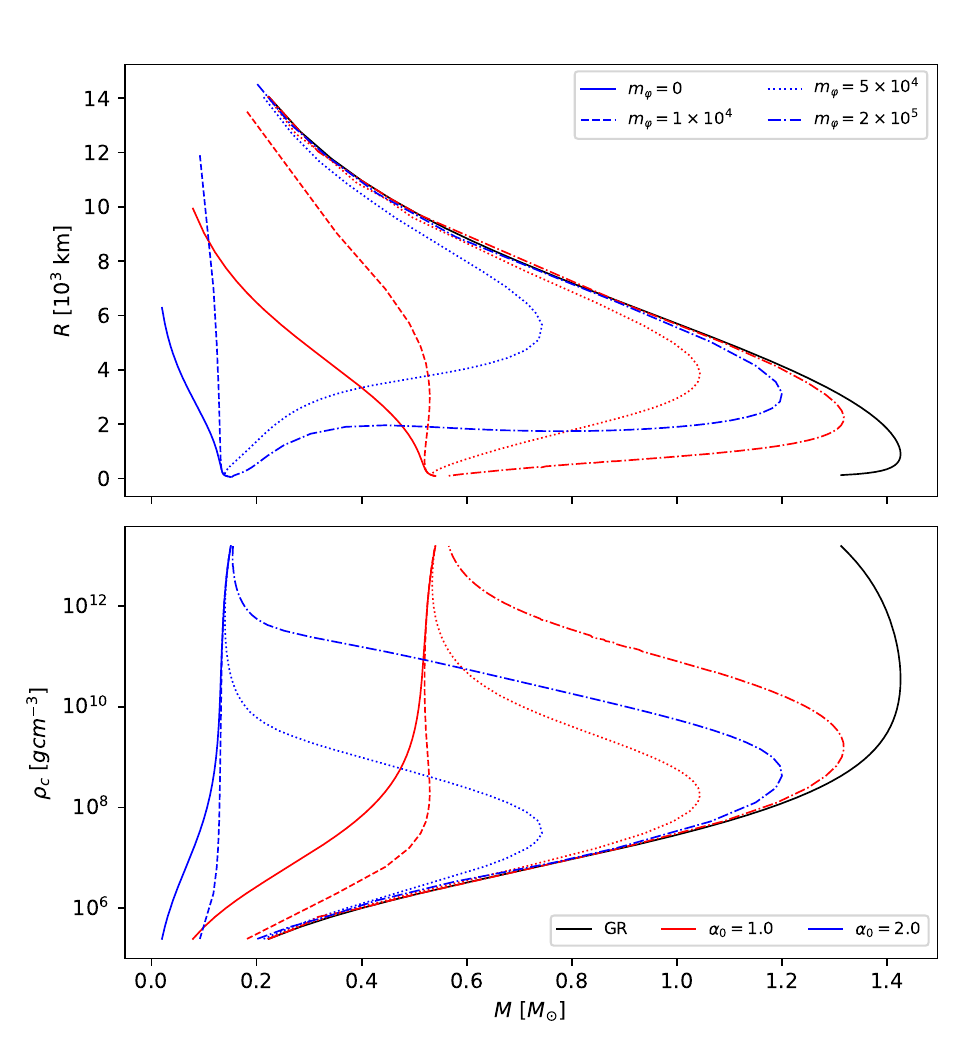}
    \caption{ The radius (upper) and central density (lower panel) as a function of mass for WDs in scalar-tensor theory. Results are given for different values of $\alpha_0$ and dimensionless scalar field mass $m_{\Tilde{\varphi}}$.}
    \label{fig:MassDensityRadius_STT}
    
    \includegraphics[width=\linewidth]{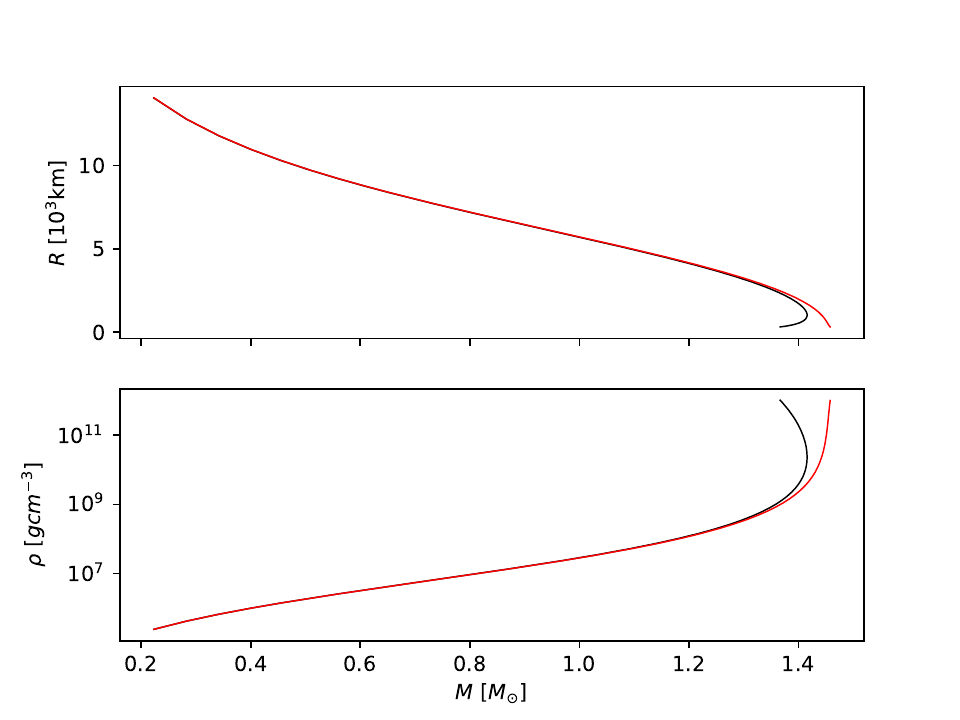}
    \caption{Mass-radius (upper) and mass-central density relation (lower panel). Each given for Newtonian gravity (red) and general relativity (black).}
    \label{fig:MassDensityRadius_NewtGR}
\end{figure}
 
\par In the following, we will begin reviewing the mass-radius relations obtained for STT, GR and its weak-field limit as well as discuss some important aspects regarding the numerical analysis. We then present all results related to the cooling process, i.e. thermal heat and crystallization, of WDs and their respective ages. The latter is subdivided into two subsections for clearer understanding: we first compare the outcomes of STT for different values of the parameters $\alpha_0, m_{\Tilde{\varphi}}$ introduced in \eqref{eq:paramspotentialSTT} and then continue to discuss the differences between GR and its Newtonian limit.

Massless STTs deviate from GR even in the weak gravitational limit. Hence, the coupling parameter $\alpha_0$ is severely constrained by various experiments in this regime \cite{PhysRevD.54.1474, antoniadis2013massive}. However, a sufficiently massive scalar field can ease this restriction, yielding a combined constraint on $\alpha_0$ and $m_{\tilde{\varphi}}$ together \cite{Perivolaropoulos:2009ak,Hohmann:2013rba,Scharer:2014kya}. In our distance range, there are virtually no constraints on $\alpha_0$ if ${m_{\Tilde{\varphi}} c^2 > 2 \times 10^{-14}eV}$ \cite{Seymour:2020yle, Alsing:2011er}. We work with $\alpha_0 = 1, 2$.

\subsection{Mass-Radius Relation}

\par The field equations \eqref{eq:scalartensorFE} describing the stellar equilibrium equations present a system of differential equations that is closed with an appropriate EoS, in this case the Chandrasekhar EoS \eqref{eq:ChandrasekharEOS}. In order to solve this system numerically, we take the following approach: 

\begin{enumerate}
    \item We work with natural units ${c, \hbar, G = 1}$ as well as ${10^nM_\odot = 1}$ and use dimensionless pressure, energy density and radial variables in the field equations. We obtain, e.g., the dimensionless pressure through 
    \begin{equation*}
        P[1]=\frac{p}{p_\mathrm{dim}} \quad\mathrm{with}\quad p_\mathrm{dim} = \frac{c^8}{(10^nM_\odot)^2G^3}\;.
    \end{equation*}
    The integer $n$ is chosen in such a way that the dimensionless central pressures considered for WDs vary around the order of $10$. Here, $n=7$.
    \item The values for $m_{\Tilde{\varphi}}$ given in the figures are dimensionless ones. They are given by 
    \begin{align*}
        m_{\Tilde{\varphi}}[1]=\frac{m_{\Tilde{\varphi}}}{m_{\Tilde{\varphi},\mathrm{dim}}} \quad\mathrm{with}\quad m_{\Tilde{\varphi}, \mathrm{dim}} = \frac{2\times10^{-5}\;c^2}{10^nM_\odot\; G}\;.
    \end{align*}
    Such that $m_{\Tilde{\varphi}}[1] \in [1\times 10^{4}, 2 \times 10^{5}]$ corresponds to $m_{\Tilde{\varphi}}c^2[eV] \in [1.3\times 10^{-13}, 2.7 \times 10^{-12}]$ and the values considered are well above the constraint imposed by observations.
    \item The numerical radial zero and infinity are set so that the obtained stellar masses vary insignificantly, less than $1\%$, when reducing or incrementing these quantities.
    \item The system to be solved in STT poses a boundary value problem, i.e. boundary conditions are given at the stellar center as well as at infinity. We implement a shooting method with $\Tilde{\varphi}_0$ as the only shooting parameter\footnote{Note that the temporal metric function $\Tilde{\phi}$ does not enter the equations \eqref{eq:scalartensorFE} directly and its derivative $\Tilde{\phi}'$ can be eliminated using its respective field equation \eqref{eq:phiSTTFE}.} to reduce it to an initial value problem. Simply put, the algorithm searches for the initial value of the scalar field at the center of the star $\Tilde{\varphi}_0$ that leads to the correct boundary condition $\Tilde{\varphi}_\infty$ at infinity when solving the system.
    \item Since considering massive models leads to stiff equations (i.e. the numerical method for solving may become unstable), fine-tuning of the parameters can be needed to obtain a result. For example adapting the initial guess for scalar field at the stellar center or/and the numerical zero as well as infinity.
\end{enumerate}


Our numerical implementation for the results in this section is based on the work done by \cite{2016PhRvD..93h4038Y}. The resulting mass-radius relation for STTs is depicted in Fig.\ \ref{fig:MassDensityRadius_STT} for different values of $m_{\Tilde{\varphi}}$ and $\alpha_0$. We observe the same qualitative behaviour as seen for neutron stars (NS). Namely, massive STTs are confined between the massless case and the GR solution for a set $\alpha_0$. However, contrary to the NS case, the maximum mass in Brans-Dicke theory is always smaller than its GR counterpart. This can be explained by considering that WDs have lower central energy densities, but might also be related to the EoS. It is also the case for NS that the STT mass-radius curves fall below the GR results towards lower central densities in \cite{2016PhRvD..93h4038Y}. This is consistent with the outcomes in $f(R)$ gravity, where assuming  $f(R)=R+\alpha R^2$, sub-Chandrasekhar mass WDs are observed for positive $\alpha$ \cite{2022PhLB..82736942K}. Further, we see that the higher $\alpha_0$ and more massive the STT, the higher is the Chandrasekhar limiting mass.

Fig.\ \ref{fig:MassDensityRadius_NewtGR} shows the obtained mass-radius as well as mass-central density relation for WDs in Newtonian gravity and GR. As expected, general relativistic effects only become relevant for high central densities, hence low stellar radii or high masses. Thus, the difference between the two theories becomes evident in this regime. We further obtain a Chandrasekhar limiting mass of $M_\mathrm{Ch}\approx1.42M_\odot$ for GR which is in good accordance with existing literature.

\begin{figure}[ht]
    \centering
    \includegraphics[width=\linewidth]{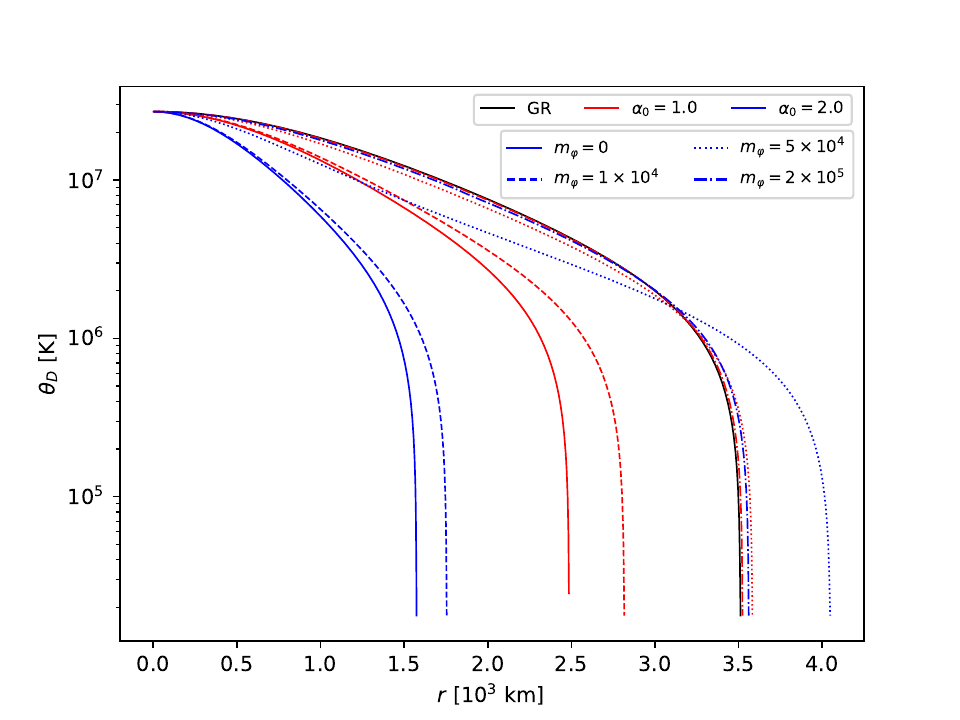}
    \caption{Debye temperature as a function of radius within a carbon WD with $p_c = 7.2 \times 10^{25} \mathrm{g}\,\mathrm{cm}^{-1}\mathrm{s}^{-2}$ in scalar-tensor theory. Results are shown for different values of $\alpha_0$ and dimensionless scalar field mass $m_{\Tilde{\varphi}}$}
    \label{fig:Debye_STT}
\end{figure}

\subsection{Crystallization and Cooling of White Dwarfs}

We now turn towards the physical processes that affect the resulting cooling age for WDs in Newtonian gravity, GR and STT. Note that the numerical analysis for this section has been conducted in cgs units, contrary to the previous section.

\subsubsection{Scalar-Tensor Theory}

\begin{figure}[ht]
    \centering
    \includegraphics[width=\linewidth]{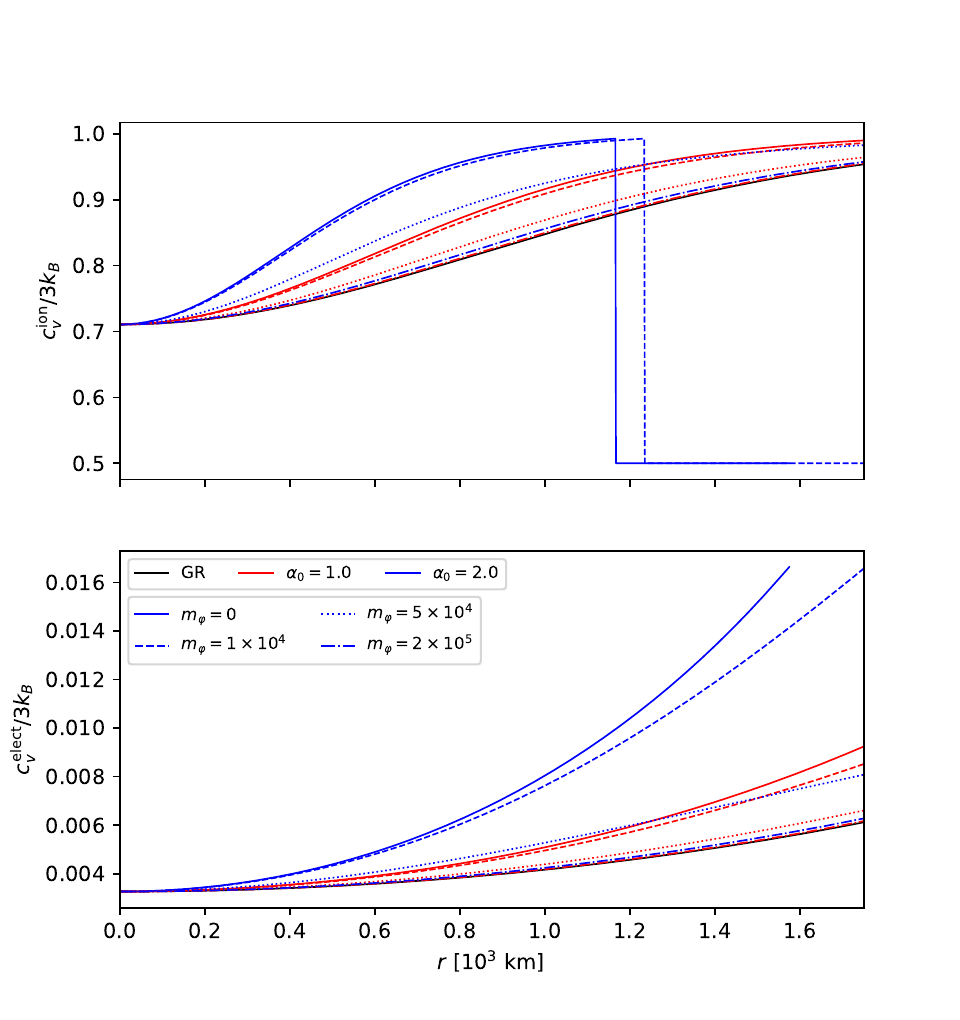}
    \caption{ Radial profile of $c_v$ inside WD with $p_c = 7.2 \times 10^{25} \mathrm{g}\mathrm{cm}^{-1}\mathrm{s}^{-2}$ in scalar-tensor gravity. Ionic specific heat $c_v^\mathrm{ion}$ and electron specific heat per ion $c_v^\mathrm{ion}$ are given at a temperature of $10^7$K for different values of $\alpha_0$ and scalar field mass $m_{\Tilde{\varphi}}$. }
    \label{fig:SpecificHeat1e7_STT}
    \includegraphics[width=\linewidth]{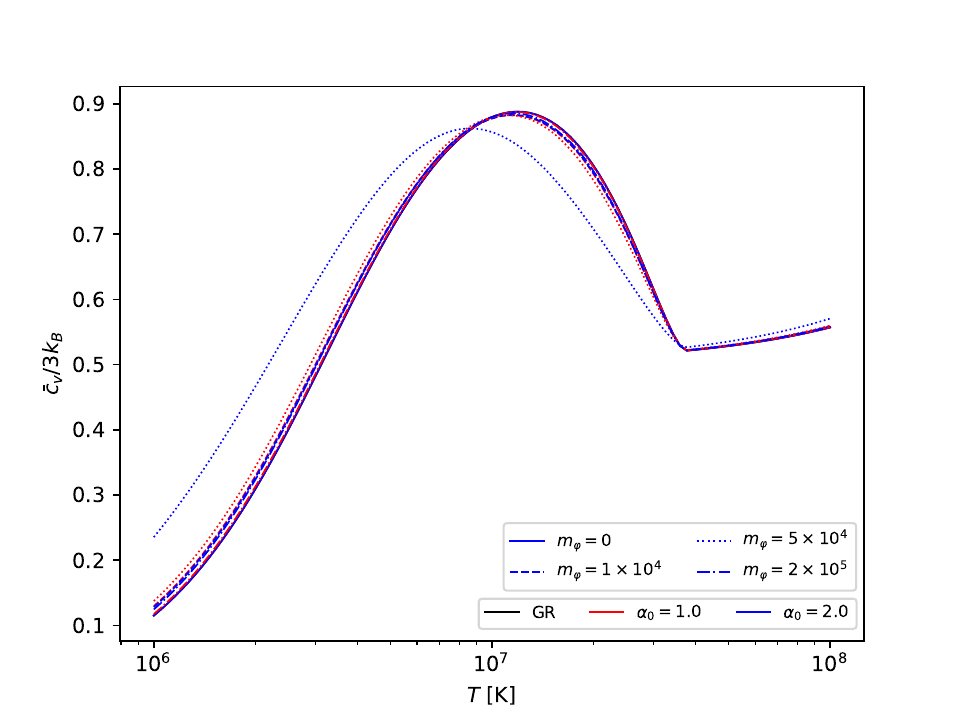}
    \caption{ Mean specific heat $\Bar{c}_v$ as a function of temperature $T$ for a carbon WD with $p_c = 7.2 \times 10^{25} \mathrm{g}\mathrm{cm}^{-1}\mathrm{s}^{-2}$ for scalar-tensor theories with different values of $\alpha_0$ and scalar field mass $m_{\Tilde{\varphi}}$.}
    \label{fig:MeanSpecificHeat_STT}
\end{figure}

We present our results for a carbon WD with a central pressure of $p_c = 7.2 \times 10^{25} \mathrm{g}\mathrm{cm}^{-1}\mathrm{s}^{-2}$ for the Debye temperature, electron and ion specific heat at $T = 10^7$K as well as mean specific heat in Figs. \ref{fig:Debye_STT}--\ref{fig:MeanSpecificHeat_STT}, respectively. While the overall shape of the Debye temperature and specific heats stays comparable for different theories, it becomes evident how the different density profiles for each theory change the radial development of these functions. Note that the sharp fall-off in the ion specific heat comes from the onset of crystallization. Instead of a constant, the ion specific heat takes the form given in \eqref{eq:ionspecificheat} when matter is crystallized. We see in Fig.\ \ref{fig:SpecificHeat1e7_STT} that the radius at which this takes place depends  on the gravity theory considered. This critical radius is set by the ratio $\Gamma$ which again depends on the density profile within the star and therefor on the gravity theory. Even though we can still recognise differences in the mean specific heat, the deviations become less dominant for most theories. We expect the large deviation for $\alpha_0=2$ and $m_{\Tilde{\varphi}}=5 \times 10^5$ to follow from the fact the stellar radius in this theory is larger than in all others at $p_c = 7.2 \times 10^{25} \mathrm{g}\mathrm{cm}^{-1}\mathrm{s}^{-2}$. This is also visible in Fig.\ \ref{fig:Debye_STT}.

\begin{figure*}
  \begin{minipage}{.48\textwidth}
    \centering
    \includegraphics[width=.95\linewidth]{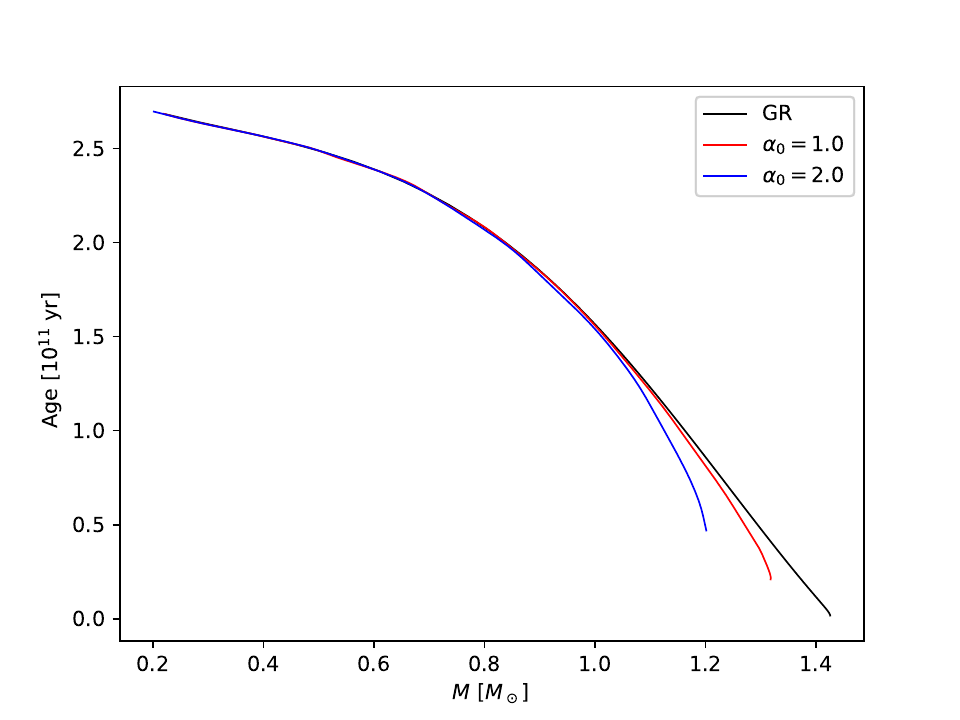}
    \caption{Age of carbon WDs as a function of their mass when they cool down from $10^8$K to $10^6$K for scalar-tensor theories with different values of $\alpha_0$ and a scalar field mass of $m_{\Tilde{\varphi}}=2 \times 10^{5}$.}
    \label{fig:CoolingAge_STT}
    \includegraphics[width=\linewidth]{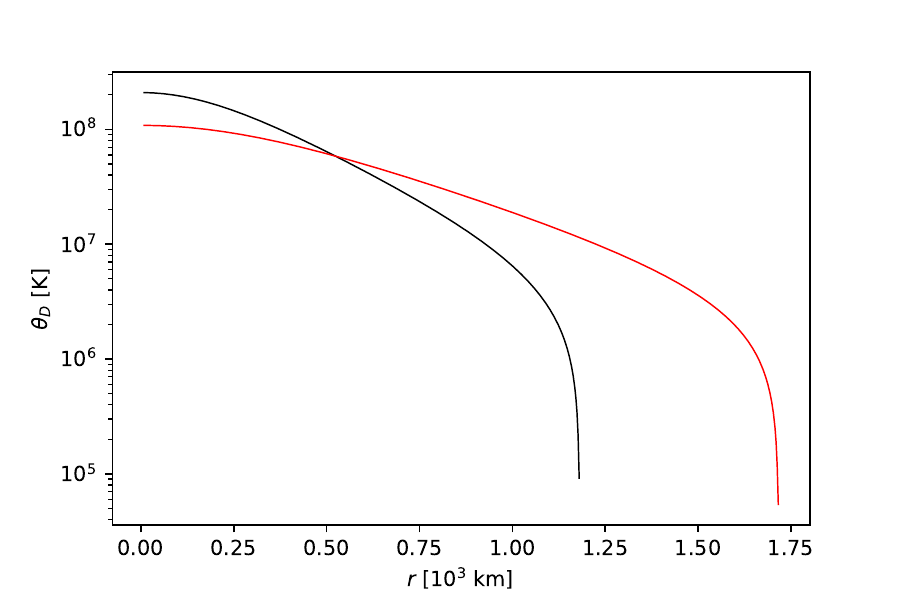}
    \caption{Debye temperature as a function of radius within a carbon WD with $M = 1.42 M_\odot$ in Newtonian gravity (red) and GR (black).}
    \label{fig:Debye_NewtGR}
  \end{minipage} \quad
  \begin{minipage}{.48\textwidth}
    \centering
    \includegraphics[width=\linewidth]{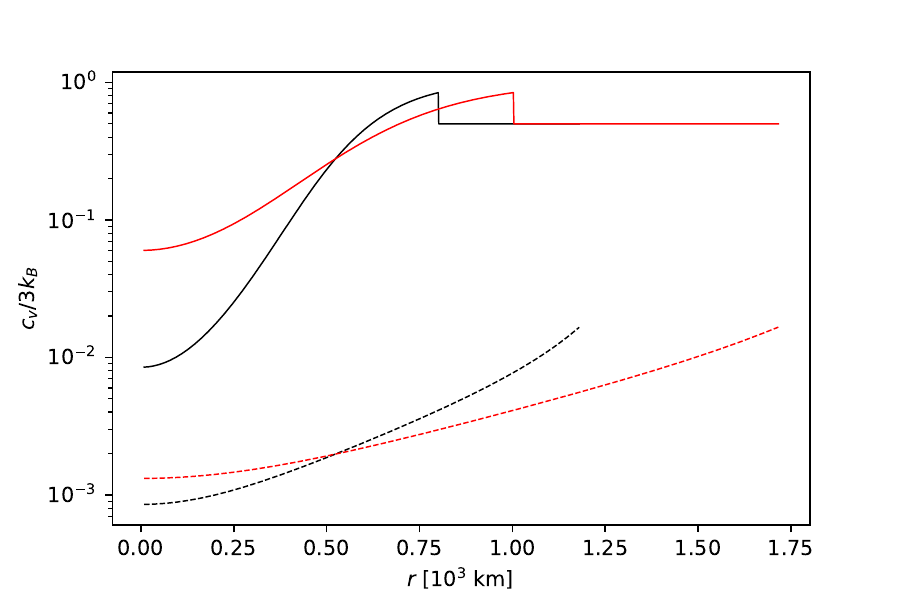}
    \caption{Radial profile of $c_v$ inside a WD with $M = 1.42 M_\odot$ for Newtonian gravity (red) and GR (black). Continuous lines represent $c_v^\mathrm{ion}$ and dashed lines $c_v^\mathrm{ion}$, both at a temperature of $10^7$K.}
    \label{fig:SpecificHeat1e7_NewtGR}
    \includegraphics[width=\linewidth]{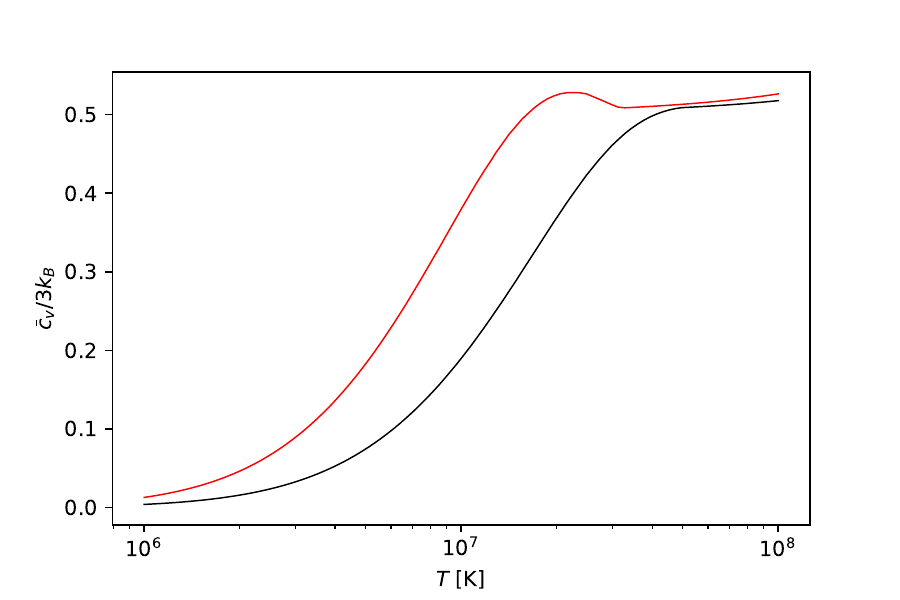}
    \caption{Mean specific heat $\Bar{c}_v$ as a function of temperature $T$ for a carbon WD with $M = 1.42 M_\odot$ in Newtonian gravity (red) and GR (black).}
    \label{fig:MeanSpecificHeat_NewtGR}
  \end{minipage}
\end{figure*}

Finally, Fig.\ \ref{fig:CoolingAge_STT} displays the cooling ages for the most massive STT with $m_{\Tilde{\varphi}}=2\times10^5$ and $\alpha_0=1, 2$ as well as GR. We observe that massive STTs can significantly reduce the cooling age of WDs at higher masses.

\subsubsection{GR and Newtonian Gravity}

Our results are illustrated for a carbon WD with $M = 1.42 M_\odot$ in Newtonian gravity and GR. Fig.\ \ref{fig:Debye_NewtGR} shows the radial profile of the Debye temperature inside a WD. We note that for the same observed mass, the stellar radius differs significantly between Newtonian gravity and GR. Thus, the radial evolution of the Debye temperature within the star is significantly distinct between the two theories. The same qualitative behaviour can be observed for the electron and ion specific heat in Fig.\ \ref{fig:SpecificHeat1e7_NewtGR} for a temperature of $T=10^7$K. These differences cause the discrepancies for the mean specific heat as a function of the temperature $T$ seen in Fig.\ \ref{fig:MeanSpecificHeat_NewtGR}.

\begin{figure}[hb]
    \centering
\end{figure}

The respective cooling ages for carbon WDs as a function of stellar mass are shown in Fig.\ \ref{fig:CoolingAge_NewtGR} for Newtonian gravity and GR. We can clearly observe that relativistic effects become non-negligible for more massive WDs while they can be safely neglected for lower mass stars. The figure further shows the significance of incorporating the crystallization process in comparison to excluding it from the cooling model.

\section{Conclusions} \label{sec:conclusions}

In this paper we have studied the effects of scalar-tensor gravity (STT) on the mass-radius relation and cooling processes of white dwarfs (WDs). In addition, we showed the relevance of considering fully relativistic theories when analyzing such stars.

\begin{figure}
    \centering
    \includegraphics[width=\linewidth]{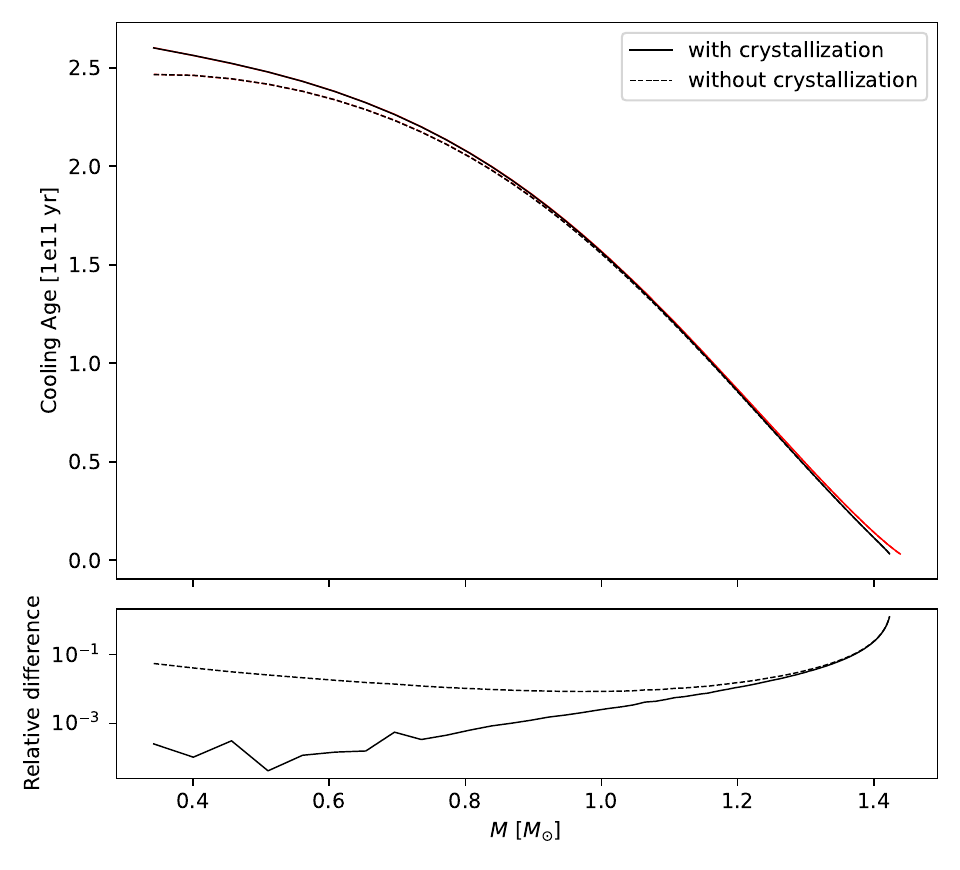}
    \caption{Age of carbon WDs as a function of their mass when they cool down from $10^8$K to $10^6$K in  Newtonian gravity (red) and GR (black). Upper panel: Results obtained by integrating \eqref{eq:totallum} and plotted considering (solid) and excluding (dashed) the crystallization process. Lower panel: relative age difference between Newtonian gravity and GR.}
    \label{fig:CoolingAge_NewtGR}
\end{figure}

Firstly, we presented the field equations in a Brans-Dicke like STT charecterised by a non-minimal coupling parameter $\alpha_0$ and an effective mass $m_{\Tilde{\varphi}}$. We described the equilibrium configuration of WDs in STT as well as shortly stated their general relativistic analogue. Hereby, we assumed that these stars can be described as a spherically symmetric, static object composed of fully degenerate electrons. Thus, matter can be described by the analytic Chandrasekhar equation of state (EoS). We saw from the resulting mass-radius relation in Fig.\ \ref{fig:MassDensityRadius_STT} that Brans-Dicke theories lead to a reduction of the maximum mass, as contrary to neutron stars (NSs) where the maximal mass is increased. This effect most likely results from the lower central densities and equation of state considered, when analyzing WDs. Furthermore, it is consistent with previous results found for $f(R)$ gravity \cite{2016PhRvD..93h4038Y}. STT could therefor potentially explain sub-Chandrasekhar WD and thereby under-luminous type Ia supernovae. Otherwise, we observe the same qualitative behaviour as for NSs in STT, a higher $\alpha_0$ leads to larger deviations from GR while a larger $m_{\Tilde{\varphi}}$ effectively suppresses the scalar field and so reduces this deviation. Further, we successfully reproduced the difference in equilibrium configurations between Newtonian gravity and general relativity (GR) in Fig.\ \ref{fig:MassDensityRadius_NewtGR} \cite{carvalho2018general}. A fully relativistic treatment becomes essential for massive WDs.

Thereafter, we continued to determine relevant quantities to obtain the cooling age of a WD star. Looking at the nondegenerate surface layers and starting from the photon diffusion equation as well as relativistic hydrostatic equilibrium equation, we derived the luminosity in dependence of the chemical composition, total mass and temperature of the star. In the stellar core, we have to take two relevant energy sources into account: the residual thermal energy of the WD progenitor and the latent heat released during the crystallization of matter. The addition of these two quantities leads to the total core luminosity. Ultimately, by equating the core and surface layer luminosity, we developed an analytic expression giving the so-called cooling age of a WD star. That is, the time it takes such a star to cool down from $10^8$K to $10^6$K. We then briefly described the equivalent surface layer equations in GR, as well as its weak-gravitational limit, and discussed how different gravity theory affects the cooling equations. We clearly saw that modified gravity theories influence the interior properties of the star, like the Debye temperature in Fig.\ \ref{fig:Debye_STT} or the specific heat in Fig.\ \ref{fig:SpecificHeat1e7_STT}. Consequently, the crystallization and cooling processes are affected. We note that STT considerably shortens the cooling time for a WD of a certain mass compared to GR as shown in Fig.\ \ref{fig:CoolingAge_STT}. On the other hand, GR increases the cooling age of massive stars with respect to Newtonian theory as seen in Fig.\ \ref{fig:CoolingAge_NewtGR}.
It is important to note that, in general, crystallization extends the overall cooling process, regardless of the model of gravity, due to the additional energy present as latent heat, which must be radiated away from the star’s surface. However, as previously discussed, STTs actually shorten the cooling process. This feature is desirable for explaining the existence of white dwarfs that appear “older than the Universe.”
This effect of “too old stars" is even more pronounced for low-mass white dwarfs, suggesting that, within the framework of Newtonian physics, one can account for extremely old white dwarfs \cite{2019ApJ...881L...3M}. None of the well-known and accepted scenarios \cite{1997ApJ...482..420L} has been sufficient to explain this unusual phenomenon. Given that modified gravity effects accelerate different phases of stellar evolution compared to Newtonian gravity \cite{Wojnar:2020frr,Benito:2021ywe,2021PhRvD.104j4058W,2023EPJC...83..492G,2023PhRvD.108b4016K}, even if modifications of GR are small, they can be significant and thereby explain at least some observed phenomena in astrophysics. 

As a final note, we should also mention that the cooling mechanism used in this study requires further refinement. The first steps toward a more realistic model would involve developing a more comprehensive model for the atmosphere that includes both the realistic equation of state and opacity, as well as a more accurate equation of state for the deep interior of WDs. Work in these areas is currently underway, and we hope to provide a more precise description of WD cooling processes, potentially incorporating observational constraints, in the near future.

\section*{Acknowledgements}
SV is grateful to the members of the theoretical physics depertment of the Universidad Complutense, Madrid, Spain for their warm welcome during working on this manuscript and the Erasmus+ traineeship programme of the University of Tartu, Estonia.
AW acknowledges financial support from MICINN (Spain) {\it Ayuda Juan de la Cierva - incorporaci\'on} 2020 No. IJC2020-044751-I and from  the Spanish Agencia Estatal de Investigaci\'on Grant No. PID2022-138607NB-I00, funded by MCIN/AEI/10.13039/501100011033, EU and ERDF A way of making Europe. 
LJ was supported by the Center of Excellence ``Foundations of the Universe'' (TK202U4) funded by the Estonian Ministry of Education and Research. 
DD acknowledges financial support via an Emmy Noether Research Group funded by the German Research Foundation (DFG) under Grant No. DO 1771/1-1. The partial support of KP-06-N62/6 from the Bulgarian science fund is also gratefully acknowledged.

\bibliographystyle{utphys}
\bibliography{biblio}

\end{document}